\documentclass[%
 reprint,
 amsmath,amssymb,
 aps,
pra,
]{revtex4-2}
\usepackage{xcolor}
\usepackage{graphicx}
\usepackage{dcolumn}
\usepackage{upgreek,xspace}
\usepackage{bm}
\usepackage[mathlines]{lineno}
\graphicspath{{Figures/}} %

\begin{document}


\title{A simulation-based training framework for machine-learning applications in ARPES}

\author{MengXing Na}
\affiliation{Quantum Matter Institute, University of British Columbia, Vancouver, British Columbia, V6T 1Z4, Canada}
\affiliation{Department of Physics \& Astronomy, University of British Columbia, Vancouver, British Columbia, V6T 1Z1, Canada}
\email{mengxing.na@ru.nl}

\author{Chris Zhou}
\affiliation{Quantum Matter Institute, University of British Columbia, Vancouver, British Columbia, V6T 1Z4, Canada}
\affiliation{Department of Physics \& Astronomy, University of British Columbia, Vancouver, British Columbia, V6T 1Z1, Canada}

\author{Sydney K. Y. Dufresne}
\affiliation{Quantum Matter Institute, University of British Columbia, Vancouver, British Columbia, V6T 1Z4, Canada}
\affiliation{Department of Physics \& Astronomy, University of British Columbia, Vancouver, British Columbia, V6T 1Z1, Canada}

\author{Matteo Michiardi}
\affiliation{Quantum Matter Institute, University of British Columbia, Vancouver, British Columbia, V6T 1Z4, Canada}
\affiliation{Department of Physics \& Astronomy, University of British Columbia, Vancouver, British Columbia, V6T 1Z1, Canada}
\email{matteo.michiardi@ubc.ca}

\author{Andrea Damascelli}
\affiliation{Quantum Matter Institute, University of British Columbia, Vancouver, British Columbia, V6T 1Z4, Canada}
\affiliation{Department of Physics \& Astronomy, University of British Columbia, Vancouver, British Columbia, V6T 1Z1, Canada}
\email{damascelli@physics.ubc.ca}

\begin{abstract}
In recent years, angle-resolved photoemission spectroscopy (ARPES) has advanced significantly in its ability to probe more observables and simultaneously generate multi-dimensional datasets. These advances present new challenges in data acquisition, processing, and analysis. Machine learning (ML) models can drastically reduce the workload of experimentalists; however, the lack of training data for ML -- and in particular deep learning -- is a significant obstacle. In this work, we introduce an open-source synthetic ARPES spectra simulator -- \textit{aurelia} -- for the purpose of generating the large datasets necessary to train ML models. As a demonstration, we train a convolutional neural network to evaluate ARPES spectra quality -- a critical task performed during the initial sample alignment phase of the experiment. We benchmark the simulation-trained model against actual experimental data and find that it can assess the spectra quality more accurately than human analysis, and swiftly identify the optimal measurement region with high precision. Thus, we establish that simulated ARPES spectra can be an effective proxy for experimental spectra in training ML models. 
\end{abstract}

\maketitle
\section{Introduction}
Our understanding of materials at the atomic and electronic levels is crucial for advancing fundamental physics and technological developments. In the study of the electronic properties of solids, angle-resolved photoemission spectroscopy (ARPES) has emerged as an indispensable tool with the unique capability to probe the momentum-resolved electronic structure~\cite{Damascelli2004, Sobota2021}. In recent years, the capabilities of ARPES have expanded to include time-domain dynamics~\cite{Boschini2024, Na2023}, spatial scanning~\cite{Cattelan2018}, spin~\cite{Okuda2017}  and orbital angular momentum~\cite{Beaulieu2020} resolution, and more. The increasing number of observables presents new challenges in both data acquisition and analysis, particularly as the acquisition time increases. Usually, the experimentalist scans the sample to find the optimal spot for measurements and monitors the spectra quality regularly. This process is time-intensive and critically dependent on the experimentalist's experience. 

A more consistent and efficient approach would be to assess the data quality using machine-learning (ML) algorithms. Integrating ML algorithms with scientific research has opened up new possibilities in data acquisition and analysis~\cite{Carleo2019}, and has laid the groundwork for AI-driven experiments~\cite{Krull2020, Maffettone2023}. These ML algorithms have the potential to dramatically speed up time-intensive procedures, thereby freeing experimentalists to take on more specialized tasks. While ARPES datasets inherently encode a suite of complex factors -- from the material's electronic properties to the experimental geometry and characteristics -- their structures are multi-dimensional matrices, which are suited to various ML techniques, including convolutional neural networks (CNNs). These CNNs have been employed successfully on physics-related data to denoise and extract features~\cite{Peng2020, Kim2021, Restrepo2022}. Using the Boltzmann machine, Yamaji et al.~\cite{Yamaji2021} uncovered peak structures in the self-energy of cuprates superconductors and Lee et al.~\cite{Lee2023} applied K-nearest neighbour and feed-forward neural networks to predict spectral functions and infer model parameters from ARPES data. 

However, these previous studies focus on the post-processing of acquired experimental data, and the ML models are trained using a small number of locally available datasets. Algorithms trained with big data from online databases such as Materials Project have also been demonstrated, though to a much smaller extent~\cite{Xi2022, Xian2023}. Training an ML model -- particularly one with sufficient reliability and versatility to handle different materials and experimental conditions -- requires tens of thousands of images and the appropriate flags and labels. However, despite efforts to streamline and standardize the collection of ARPES data and metadata, a common database does not yet exist. The lack of suitable datasets is a significant obstacle to training and implementing ML algorithms in ARPES data acquisition and processing~\cite{Alzubaidi2023}. While this highlights the need for collaborative efforts to create shared data, such efforts take time to come to fruition.

\begin{figure*}[ht!]
    \centering
    \includegraphics{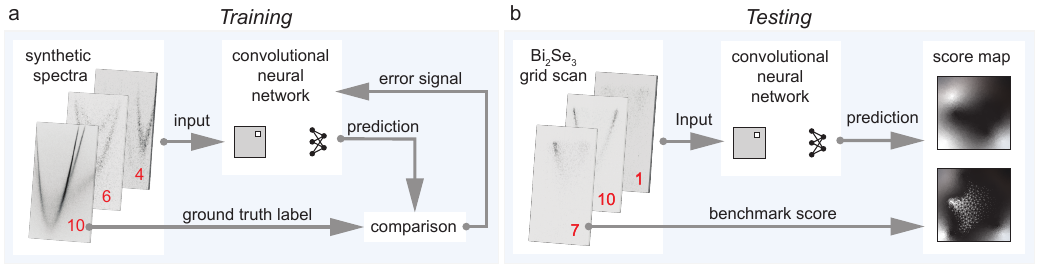}
    \caption{\textbf{A synthetically-trained convolutional neural network.} \textbf{(a)} The training process: Synthetic data produced by the simulator are given to the CNN along with the ground truth label, which is a quality assessment score calculated from simulation parameters. \textbf{(b)} The trained model is tested with experimental spectra -- specifically a grid scan across the cleaved surface of a Bi$_2$Se$_3$ sample. The machine-learning prediction gives a score (MLS) for each spectrum, which results in a score map of the sample surface. For benchmarking purposes, a score made through analysis (BMS) is generated and compared with the MLS.} 
    \label{Fig: procedure}
\end{figure*}

One approach to circumvent the lack of experimental data is to use synthetic data to train ML models~\cite{Sun2022, Chen2024}. In this paper, we present a general ARPES data simulator named \textit{aurelia} that mimics experimental data for training ML models~\cite{NaGit2024}. While traditional ARPES simulations aim to accurately describe a material's electronic structure as measured under ideal experimental conditions, our approach is to generate diverse ARPES spectra under a variety of experimental conditions. As an excercise, we tackle the task of assessing data quality and determining the best region to probe. The procedure for training and testing is illustrated in Fig.\,\ref{Fig: procedure} a and b, respectively. In Sec.\,\ref{Sec: Sim} we show how the simulator can produce realistic ARPES datasets. These synthetic spectra are given to the CNN as inputs. In Sec\,\ref{Sec: QS}, we define the \textit{ground truth} labels, which -- for the purposes of our spectra assessment task -- is a quality score index. The details of the ML algorithm is given in Sec.\,\ref{Sec: ML}.

Once trained, we benchmark the effectiveness of ML model with experimental data (Sec.\,\ref{Sec: test}). Using a newly developed $\upmu$-ARPES setup~\cite{Dufresne2024}, we perform real-space grid scans across the cleaved surface of Bi$_2$Se$_3$. The spectrum at each point of the grid is assessed by the CNN, resulting in a \textit{machine-learning score} (MLS) map. Since the experimental Bi$_2$Se$_3$ spectra do not have ground truth labels, we instead create a \textit{benchmark score} (BMS) through analysis. Lastly, we compare the MLS and BMS and show that the MLS is more sensitive to variations in the spectra and can more accurately determine the optimal measurement location. While we demonstrate the ML model on Bi$_2$Se$_3$ spectra, the CNN is trained on spectra from many different dispersions. Therefore, our ML model can assess spectra from any sample. This result is an essential step towards AI-driven ARPES experiments. The simulator \textit{aurelia} is open source and versatile, and can be designed to train ML algorithms toward a variety of tasks~\cite{NaGit2024}. 
\begin{figure}[hb!]
    \centering
    \includegraphics{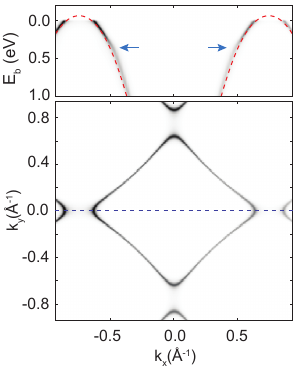}
    \caption{\textbf{Sample of the simulated photoemission intensity $I(\mathbf{k},\omega)$ calculated on a square lattice}. The red dashed lines show the bare electronic structure calculated from tight-binding model (Eq.\,\ref{Eq: tb}), $E_0 = -1.14$, $t = -0.68$, $r = 0.23$ and $a = 4.19$. The dispersion at $k_y=0$ is shown on the top row, while the Fermi surface is shown on the bottom. A small electron-boson kink indicated by blue arrows using Eq.\,\ref{Eq: SE}, where $R = 0.007$, $\omega_0=0.3$, $\gamma = 21.4$, $\Sigma_0 = 0.02$ and  $\Sigma_1 = 0.093$. The photoemission matrix elements are modelled using a smooth curve that modulates the photoemission intensity.}
    \label{Fig: specfun}
\end{figure}

\begin{figure*}
    \centering
    \includegraphics{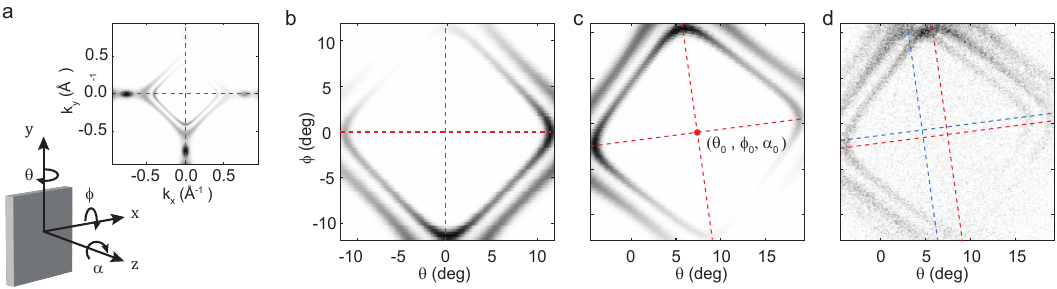}
    \caption{\textbf{Momentum-to-angle conversion and simulating counting statistics.} \textbf{(a)} Definition of rotation angles and the ARPES intensity calculated from a square tight-binding model in momentum space. \textbf{(b)} Momentum-to-angle conversion for zero offset-angles, interpolated over a square grid of angles. \textbf{(c)} The momentum-to-angle conversion for finite offset angles $(\theta_0, \phi_0, \alpha_0)$, as indicated by the red marker, while red dashed lines indicate the orientation. \textbf{(d)} Simulated counting statistics using the momentum-to-angle conversion shown in panel c: a flake domain at different offset angles is added and indicated by blue dashed lines, while red dashed lines indicate the dominant (original) domain; lastly, we add a flat background and increase the detector response in the center of the image.}
    \label{Fig: arpes}
\end{figure*}

\section{The \textit{aurelia} simulator}
\label{Sec: Sim}
The \textit{aurelia} simulator generates large numbers of synthetic ARPES spectra for the purpose of training supervised machine-learning algorithms. We begin with the physical parameters of the sample, which define the electronic structure and self-energy. The dispersion is simulated using a tight-binding model of s-orbitals on a lattice, including nearest and next-nearest neighbour terms. The simulation includes hexagonal, honeycomb, square, and rectangular lattices. The number of bands, hopping parameters, and on-site energy are parameters which we vary to generate an assortment of band structures. For example, in Fig.\,\ref{Fig: specfun}, the band structure is calculated using:
\begin{equation}
\begin{split}
    E(\mathbf{k})=E_0&+2t[\cos(k_x a)+\cos(k_y a)]\\
    &+rt[\cos(2k_x a)+\cos(2k_y a)].
    \label{Eq: tb}
\end{split}    
\end{equation}
Next, we calculate the spectral function $A(\mathbf{k}, \omega)$, which is given by:
\begin{equation}
    \begin{split}
        A(\mathbf{k},\omega)=\frac{-1}{\pi}\sum_m \frac{\Sigma_m''(\omega)}{[\omega-\epsilon_{\mathbf{k},m}-\Sigma'_m(\omega)]^2+\Sigma''_m(\omega)^2},\\
    \end{split}
\end{equation}
where $m$ is the band index. The linewidth and renormalization of the dispersion are given by the electronic self-energy $\Sigma = \Sigma'+i\Sigma''$. In Fig.\,\ref{Fig: specfun}, we include an electron-boson kink (indicated by blue arrows), which we define phenomenologically at an energy $\Omega$ below the Fermi level with an amplitude that captures the coupling strength, that is:
\begin{equation}
    \begin{split}
        \Sigma' &=R\left[\frac{\gamma}{(\omega+\omega_0)^2+\gamma^2}-\frac{\gamma}{(\omega-\omega_0)^2+\gamma^2}\right]\\
        \Sigma''&=\Sigma_1-\frac{\Sigma_1-\Sigma_0}{1+e^{-(\omega+\omega_0)\gamma}}.
    \end{split}
    \label{Eq: SE}
\end{equation}
The photoemission intensity is given by:
\begin{equation}
    I(\mathbf{k},\omega)=|M_\mathbf{k}^{f,i}|^2 A(\mathbf{k},\omega) f(\omega, T),
\end{equation}
where $|M_\mathbf{k}^{f,i}|^2$ is the photoemission matrix element and $f(\omega, T)$ is the Fermi-Dirac distribution at finite temperature $T$, and $A(\mathbf{k}, \omega)$ is the spectral function. The photoemission matrix element can strongly modulate the intensity, even entirely suppressing photoemission from certain bands. The matrix element also strongly depends on the experimental geometry, polarization and photon energy of light, which can be used to select bands with specific orbital character or angular momentum. While modelling of this has advanced significantly~\cite{Day2019}, it remains challenging to capture the effects of the matrix elements completely. In our simulation, we forgo accurate modelling of this term; instead, we treat the matrix element as a random modulation of the photoemission intensity with symmetry from the underlying lattice.

To calculate the photoemission intensity, we first convert the electron momenta into emission angle. We define the electron analyzer's optical axis as $\hat{z}$. The axis that is parallel (perpendicular) to the slit of the hemispherical analyzer is denoted $\hat{y}$ ($\hat{x}$). We define the polar angle $\theta$ as a rotation about the $\hat{y}$ axis, the tilt angle $\phi$ as a rotation about the $\hat{x}$ axis, and the azimuth angle $\alpha$ as about the $\hat{z}$ axis. In this step, we define the photon energy $h\nu$ and the offset angles between the vector normal to the sample surface $\hat{n}$ and the electron analyzer's optical axis. In Fig.\,\ref{Fig: arpes}a, an example of ARPES intensity at the Fermi energy (i.e. Fermi surface) is provided. In this step, the spectrum is convolved with a Gaussian function defining the detector's energy and momentum resolution ($\Delta E$, $\Delta \theta$). In Fig.\,\ref{Fig: arpes}b, the Fermi surface is interpolated over an angular grid and is shown for a range of $\pm 15^\circ$ in $\theta$ and $\phi$. Here, some grid artifacts from the interpolation are visible for sparse grids. The added effect of the offset angles $\theta_0$, $\phi_0$ and $\alpha_0$ is shown in Fig.\,\ref{Fig: arpes}c.

\begin{figure*}
    \centering
    \includegraphics{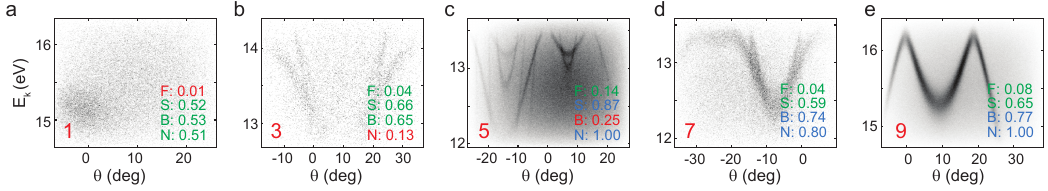}
    \caption{\textbf{Scoring the quality of simulated ARPES spectra.} The scores on the right side of the image refer to feature (F), sharpness (S), signal-to-background (B), and number of electrons collected (N). The component scores are colour-coded such that high scores ($>0.7$) are blue, low scores ($< 0.3$) are red, and middling scores are green. The total score calculated from Eq.\,\ref{Eq: total score} is given at the bottom left of each panel. The weighting factors used here are $w^S=0.6$, $w^B=1$, and $w^N=0.8$.}
    \label{Fig: QS}
\end{figure*}
We can normalize the simulated intensity in Fig.\,\ref{Fig: arpes}c [$I(E_k, \theta, \phi, \theta_0, \phi_0, \alpha_0)$] and use it as a probability distribution function $P_\text{ARPES}$. To simulate spurious ``domains", we can add the intensity from different offset angles [$P^i_\text{ARPES} = P(E_k, \theta, \phi, \theta_i, \phi_i, \alpha_0)$]. Lastly, we add the background and detector response. The background combines a constant, a polynomial, and a Shirley background, with respective amplitudes~\cite{Castle2001}. The detector response is chosen between ``flat", which is a constant, or ``center", which gives higher counts in the middle of the image, sometimes encountered in multiplexing ARPES spectrometers. The total probability distribution is given by:
\begin{equation}
    P_\text{tot} = A_\text{norm}R_\text{det}\left(\sum_i P^i_\text{ARPES} + P_\text{bkgd}\right),
\end{equation}
where $A_\text{norm}$ is a normalization prefactor. Finally, using $P_\text{tot}$, we generate $N_e$ discrete random electron events. The result is shown in Fig.\,\ref{Fig: arpes}d, where the red and blue dashed lines mark the presence of two domains of similar intensity.

The difference between Fig.\,\ref{Fig: arpes}c and Fig.\,\ref{Fig: arpes}d highlights how drastically experimental artifacts can affect the measured spectra. In creating a generalized set of spectra to train a ML model that can handle any sample measuring environment, all of the aforementioned variables -- from the tight-binding dispersion $E_{m,\mathbf{k}}$, to the electron self-energy $\Sigma$, photoemission matrix elements $|M_\mathbf{k}^{f,i}|$, temperature $T_e$, experimental resolution $(\Delta E, \Delta\theta)$, offset angles $(\theta_0, \phi_0, \alpha_0)$, background amplitudes $A_\text{flat}$, $A_\text{poly}$, $A_\text{Shir}$ and detector responsivity $R_\text{det}$ -- are randomly chosen between reasonable limits. For now, we forgo the creation of flake or rotational domains, as domain recognition is a task better handled by a separate ML model. 

\section{Ground truth labels:\\ Quality scores}
\label{Sec: QS}
In this section, we create the appropriate quality score (QS) labels for the synthetic spectra with the aim of reproducing human assessment as accurately as possible. The way to do this is by no means unique. We start by scoring each spectrum with a maximum score of $\text{s}_\text{tot} = 10$; then, as we check a list of criteria, incremental penalties $\text{s}_\text{p}$ and bonuses $\text{s}_\text{b}$ are given. 

First, we check whether a feature is present in the energy-angle window measured. The feature score is calculated by
\begin{equation}
    F=\frac{\sum_{\theta, E_k}\frac{\partial ^2|I(E_k, \theta)|}{\partial E_k\partial \theta}}{\sum_{\theta, E_k}I(E_k, \theta)}.
\end{equation}
This equation checks the intensity variation in the ARPES spectra and normalizes it to the total intensity. If there are little to no spectral features, then $F< 0.01$, and a large penalty of 4 points is incurred. Provided that there is a feature in the spectra, we consider three specific criteria for the quality assessment: 
\begin{enumerate}
    \item The sharpness of the bands (S).
    \item The signal-to-background ratio (B).
    \item The number of electrons collected (N).
\end{enumerate}
The sharpness of the bands is evaluated using the energy resolution broadening $\Delta E$, the thermal broadening $\text{k}_\text{B}T$, and the imaginary part of the self-energy $\Sigma''$. These quantities are directly available from the simulation:
\begin{equation}
    S = 1-\sqrt{\Delta E^2+(\text{k}_\text{B} T)^2+\text{min}(\Sigma'')^2}.
\end{equation}
Here, the sharpness is scored on the interval $[0,1]$, with $1$ indicating the sharpest and 0 indicating the broadest features. The penalties given are as follows:
\begin{equation}
    \begin{split}
    S &\in [1.0, 0.9], \quad\quad \text{s}_\text{p} = 0,\\
    S &\in (0.9, 0.7], \quad\quad \text{s}_\text{p} = 1,\\
    S &\in (0.7, 0.5], \quad\quad \text{s}_\text{p} = 2,\\
    S &\in (0.5, 0.3], \quad\quad \text{s}_\text{p} = 3,\\
    S &\in (0.3, 0.1], \quad\quad \text{s}_\text{p} = 5,\\
    S &\in (0.1, 0.0], \quad\quad \text{s}_\text{p} = 8.    
    \end{split}
    \label{Eq: Penalty}
\end{equation}
Next, the signal-to-background can be calculated from the ratio of the maximum image intensity with and without the spectral feature. We define the background score as follows:
\begin{equation}
    B = 1-\frac{\text{max}[R_\text{det}(I_\text{bkgd})]}{\text{max}[R_\text{det}(I_\text{bkgd}+I_\text{ARPES})]}.
\end{equation}
where $I_\text{bkgd}$ is the intensity of the background, $I_\text{ARPES}$ is the intensity of the spectral feature and $R_\text{det}$ is the responsivity of the detector. As before, $B$ is scored on the interval $[0,1]$, and the penalties are given using the same intervals defined in Eq.\,\ref{Eq: Penalty}.
\begin{figure*}
    \centering
    \includegraphics{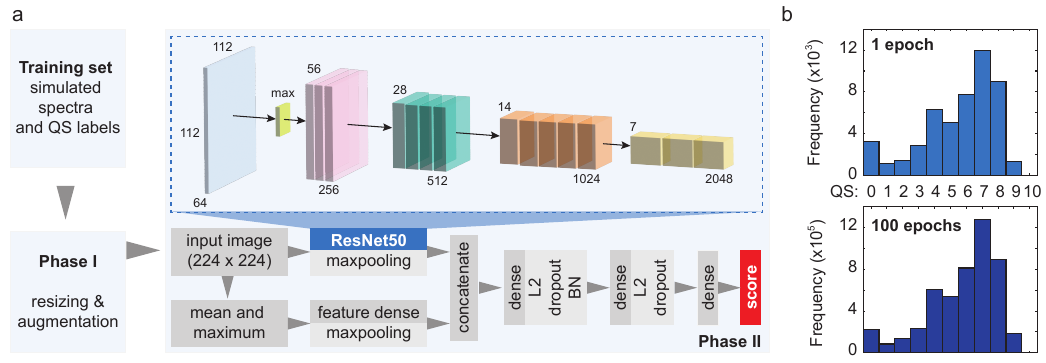}
    \caption{\textbf{Overview of the machine learning workflow.} \textbf{(a)} Schematic illustration of the deep learning pipeline for ARPES quality score prediction. In Phase I, synthetic ARPES spectra are dynamically generated using \textit{aurelia}, resized to $224\times224$ pixels, and augmented to enhance dataset diversity. Phase II consists of a dual-branch convolutional neural network (CNN), integrating a ResNet50-based image-processing pathway and a separate feature-processing branch for two intensity-based parameters (mean and maximum intensities). Feature extraction through the ResNet50 backbone involves sequential convolutional and pooling layers, reducing spatial dimensions and increasing feature depth as indicated. The resulting feature maps are concatenated with the numerical intensity features and passed through fully connected layers with L2 regularization, dropout, and batch normalization (BN), producing a numerical prediction. \textbf{(b)} Histograms show the distributions of generated QS labels for one epoch (50,000 samples) and accumulated labels over 100 epochs (5 million samples).
}
    \label{Fig: ML}
\end{figure*}

Lastly, we check the electron count $N$. As we simulate $N_\text{e}$ electron events based on a random distribution, we define:
\begin{equation}
    N = N_\text{e}/N_\text{max},
\end{equation}
where $N_\text{max} = 10^5$ is a limit corresponding to the maximum score of $N = 1$. As before, $N$ is scored on the interval $[0,1]$ and given the penalties as in Eq.\,\ref{Eq: Penalty}. 

With only penalties, we cannot fully capture human assessment. These criteria -- sharpness, signal-to-background, and electron count rate -- are interconnected in how they affect the experimentalist's decision-making. For instance, if the electron count rate is high, we save acquisition time, compensating for a poorer signal-to-background ratio. Or, if the electron count rate is poor but the signal-to-background is excellent, one can obtain high-quality spectra if the acquisition time is invested. To account for these two scenarios, we award bonuses to the $N$ and $B$ components as follows: 
\begin{equation}
    \begin{split}
        N &= 1.0, \quad \text{s}_\text{b} = 1, \\
        B &> 0.9, \quad \text{s}_\text{b} = 1,\\
    \end{split}
    \label{Eq: bonuses}
\end{equation}
On the other hand, if the spectral feature exists and is very sharp but the electron count rate and signal-to-background are relatively poor, the simulator may give a higher quality score than a human user, who (by visual inspection) cannot identify the existence of a sharp band. To account for this, we give an additional deduction:
\begin{equation}
    \begin{split}
        B+N&<0.5, \quad \text{s}_\text{c} = 2.\\
    \end{split}
    \label{Eq: deduction}
\end{equation}
Then the total score is calculated as:
\begin{equation}
    \text{S}_\text{tot} = 10 - \sum_{i = \text{S}, \text{B}, \text{N}} w^i\text{s}_\text{p}^{i} + \sum_{j = \text{N}, \text{B}}s_\text{b}^j-s_\text{c},
    \label{Eq: total score}
\end{equation}
where $s_\text{p}^i$ and $s_\text{b}^j$ are the penalty and bonus points assigned to each criterion as defined in Eq.\,\ref{Eq: Penalty} and Eq.\,\ref{Eq: bonuses}, $s_\text{c}$ is the $B+N$ combined deduction defined in Eq.\,\ref{Eq: deduction}, and $w^i$ are the user-defined weights on the interval $[0,1]$ that can be used to emphasize one criterion over another. With this, we assign quality score labels to the simulated training dataset that qualitatively agree with a human assessment.

A sample of simulations generated for training is shown in Fig.\,\ref{Fig: QS} with scores in increasing order. We can see in Fig.\,\ref{Fig: QS}a how the lack of a feature inside the image drastically lowers the score. In Fig.\,\ref{Fig: QS}b, we see how the low scoring of a single component $N$ lowers the total score. In Fig.\,\ref{Fig: QS}c, the score is 5, which points to some components scoring well -- such as $N$ and $S$-- and others less so. On the higher end of the scoring (Fig.\,\ref{Fig: QS}e), at least two criteria must score well, and some bonus points are included. Finally, we note that this algorithm-derived score relies on simulation parameters and can only be assigned to synthetic spectra. 

\begin{figure*}
    \centering
    \includegraphics{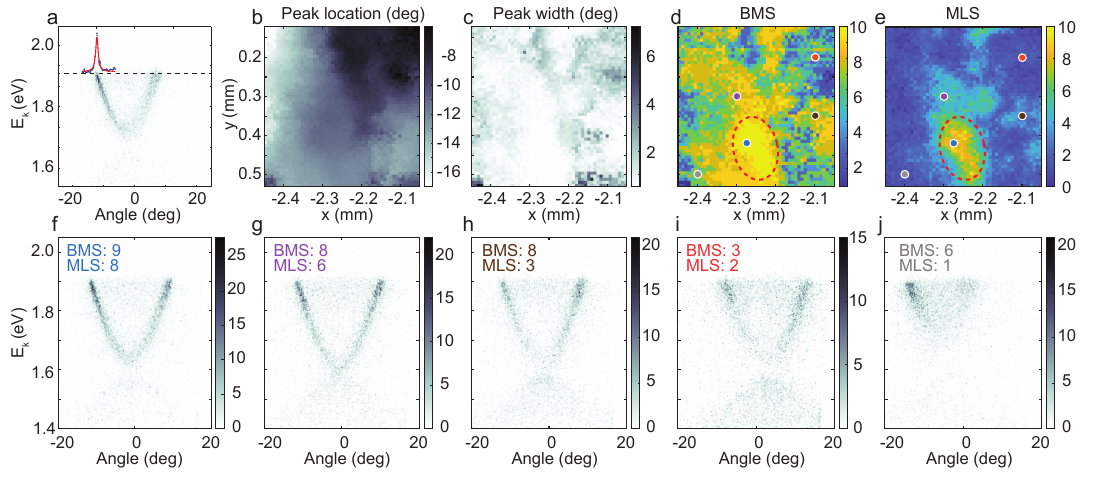}
    \caption{\textbf{Scoring the quality of experimental ARPES spectra.} \textbf{(a)} Example of the ARPES spectra measured on Bi$_2$Se$_3$ and Lorentzian fit of the momentum distribution curve (MDC) obtained by integrating the intensity in the region [1.88, 1.9]~eV for the left hand side branch of the topological surface state. \textbf{(b)} MDC peak location in angle along the ARPES cut from the fit in (a), plotted as a function of spatial position on the cleaved sample surface; different regions correspond to slightly misaligned domains. \textbf{(c)} MDC width in degrees obtained from the fit in (a); the dark regions outline the boundaries of the domains. \textbf{(d, e)} Benchmark score (BMS) of each spectrum as assessed by analysis of the TSS (Eq.\,\ref{Eq: BM score 1} and \ref{Eq: BM score 2}). \textbf{(e)} Quality score assessed by the machine-learning model. The highest scoring region is shown by dashed red lines. \textbf{(f-j)} The Bi$_2$Se$_3$ spectra at five locations on the sample; the benchmark score (BMS) and machine-learning score (MLS) are at the top of each panel.}
    \label{Fig: GS}
\end{figure*}
\section{The ML-algorithm}
\label{Sec: ML}
The deep learning model employed in this work is a modified convolutional neural network (CNN) based on the ResNet50 architecture, designed to process \textit{aurelia}-generated ARPES spectra and extract meaningful numerical scores that quantify spectra quality. As detailed in Fig.\,\ref{Fig: ML}a, the model consists of two primary input branches: an image processing pathway using ResNet50 pre-trained on ImageNet, and a secondary feature extraction pathway handling two intensity-related parameters (maximum and mean intensity values). The ResNet50 branch processes spectral images resized to $224\times 224$~pixels, followed by global average pooling, generating a feature vector of 2048 dimensions. The numerical-feature branch consists of fully connected layers: a 128-neuron layer followed by batch normalization, dropout (0.3), and ReLU activation, and a second dense layer with 64 neurons, dropout (0.3) and ReLU. These two branches are concatenated and passed through additional dense layers (256 and 128 neurons) with batch normalization, dropout, and L2 regularization to mitigate overfitting.

The training dataset (Fig.\,\ref{Fig: ML}b) consists of dynamically generated 50,000 image-label pairs per epoch. Simulation parameters -- including the definition of the band structure, the experimental geometry, detector sensitivity, and more -- are sampled from continuous uniform distributions over a reasonable intervals to reflect realistic experimental variability. To avoid training bias due to a uneven QS imbalance, we implemented an adaptive ``use-or-regenerate" sampling strategy. During the data generation, each newly simulated data point was selectively accepted, bypassed, or regenerated depending on whether its QS fell within an overrepresented or underrepresented region of the score distribution. We note, however, that the resulting distribution of QS is not uniform. We observe in Fig.\,\ref{Fig: ML}b that there are more images with $\text{QS}\in[4,8]$, less with $\text{QS}\in[0,3]$, and nearly none with $\text{QS}=10$. This is not necessarily a drawback, as the lowest scores typically correspond to spectra without discernible features, and the highest scores are rare in both simulation and experiment. Since our goal is to train a model that can distinguish subtle differences between spectra of moderate quality -- differences that may elude a human operator -- this imbalance favours the more relevant QS range.

To enhance model robustness, data augmentation and pre-processing techniques were applied to the training dataset. Augmentation included random speckles, width and height shifts, nonlinear contrast transformations, and horizontal flips. Whitening transformations were employed to normalize intensity variations across datasets. Each epoch uses a newly generated dataset to ensure the model learns from diverse variations and prevent memorization. Training was run for up to 100 epochs with early stopping triggered after 10 consecutive epochs without improvement in validation loss. Each epoch consists of 781 gradient update steps (batch size = 64). The Adam optimizer was used with an initial learning rate of 0.001, reduced adaptively via ReduceLROnPlateau (patience = 5 epochs, reduction factor = 0.2), ensuring stable convergence.

Cross-validation was performed by reserving 1,000 image-label pairs for validation. The model was trained using mean squared error (MSE) loss to predict continuous numerical scores. The final model achieved a training mean absolute error (MAE) of 0.3569 and a validation MAE of 0.3762 using dynamically generated data. For comparison, training on a fixed dataset of 50,000 samples over all epochs resulted in higher MAEs (1.083 training, 1.130 validation), confirming the advantage of dynamic data generation. These results demonstrate that our simulation-trained model generalizes well and provides accurate, automated assessment of ARPES spectra quality. 

\section{Testing data: Bi$_2$Se$_3$}
\label{Sec: test}
We assess the performance of the ML algorithm trained exclusively using the output of our simulator using experimental ARPES spectra obtained from a grid scan of a bulk cleaved sample of a topological insulator Bi$_2$Se$_3$. Here, we cannot calculate scores as shown in Sec.\,\ref{Sec: QS}, as there are no simulation parameters. However, we can analyze the experimental spectra for the sharpness, signal-to-background, and number of electrons criteria and calculate a benchmark score (BMS) to compare with the machine-learning prediction.

An example of the spectra is shown in Fig.\,\ref{Fig: GS}a. We fit the momentum-distribution curve (MDC) at the Fermi level $E_\text{F}$ between the angles [-17, -6] degrees with a Lorentzian to determine the location along the ARPES cut in kinetic energy versus angle and the width of the peak. Plotting the peak location in degrees (Fig.\,\ref{Fig: GS}b) from the fit allows us to distinguish domains with different orientation. The spatial variation of the width of the peak (Fig.\,\ref{Fig: GS}c) highlights the boundary of these domains. We use the full-width-at-half-maximum $\sigma_\text{FWHM}$ of the Lorentzian fit to determine the sharpness $S$, normalizing over the range of the dataset, so that:
\begin{equation}
    S = 1- \frac{\sigma_\text{FWHM}}{\text{max}(\sigma_\text{FWHM})}.
    \label{Eq: BM score 1}
\end{equation}
To calculate the background score $B$, we first estimate the signal-to-background ratio of the ARPES spectra. The background intensity ($I_\text{bkgd}$) is obtained by summing the intensities of all pixels with values below a fixed threshold $I_\text{th}$. Here, we choose $I_\text{th} = 1$ to exclude spectral features of Bi$_2$Se$_3$ and isolate the low-intensity background  (see Fig.\,\ref{Fig: GS}f-j). The total intensity $I_\text{total}$ is defined as the sum of intensities over all pixels in the ARPES image. The background score $B$ is then given by:
\begin{equation}
B = 1 - \frac{I_\text{bkgd}}{I_\text{total}}
\label{Eq: BM score 2}
\end{equation}
The electron count score $N$ is defined as:
\begin{equation}
N = \frac{I_\text{total}}{N_\text{max}},
\end{equation}
where $N_\text{max}$ defines the intensity limit where $N=1$. 

These three scores used to inform the penalties incurred and the final benchmarking score (BMS) following the procedure we defined in Sec.\,\ref{Sec: QS}. The BMS is shown in Fig.\,\ref{Fig: GS}d; we observe that a region approximately 100~$\upmu$m wide is shown to be the optimal spot to measure, as indicated by red dashed lines. The score predicted by the machine-learning model (MLS) is shown in Fig.\,\ref{Fig: GS}e, normalized over the dataset such that the poorest image quality is zero and the highest is 10. We see that similar regions of high scores are identified at the bottom center of the image. However, the MLS is much more precise with determining the optimal measurement region, and shows a much better contrast among high scores. This is remarkable considering that the ML-model is not explicitly trained in the recognition of flake domains; rather, the domains here manifest as a broadening of the spectral features, which is assessed by the ML model under the ``sharpness" criteria. 

In Fig.\,\ref{Fig: GS}f-j, we sample five spectra in the grid scan and compare the evaluation of the BMS and MLS. While we see a progressive increase in the spectra quality with the increase in scores, the BMS and MLS do not agree precisely. In fact, the ML model is more accurate and discerning. For instance, the BMS assigns a score of 8 for panels Fig.\,\ref{Fig: GS}g and h, while we clearly see that Fig.\,\ref{Fig: GS}g has a higher signal-to-background and sharper features. The MLS correctly differentiates the two panels and assigns the former a score of 6 and the latter a score of 3. Finally, the BMS assigns a score of 6 to Fig.\,\ref{Fig: GS}j, while the MLS gives a score of 1. Again we can visually confirm that the MLS is more accurate. This large discrepancy can be attributed to the analysis used to calculate the BMS. Specifically, the analysis only considers the left branch of the topological surface state. Indeed, the BMS map shown in Fig.\,\ref{Fig: GS}d looks remarkably like the fitted FWHM in Fig.\,\ref{Fig: GS}c. In contrast, the machine-learning model assesses the entire spectrum and is, as a result, more human-like in its assessment of the spectra quality. 

\section{Conclusion}
In this work, we have developed a simulator for ARPES spectra to generate large training datasets for machine learning. We show that these simulated spectra are a suitable proxy for experimental data in training ML models. Since these simulations are calculated analytically, we possess all the necessary parameters and metadata to be used as ground-truth labels in supervised learning. By generating synthetic ARPES spectra, fixing some parameters while randomizing others, one can train ML algorithms for various specific tasks.

Here, we have demonstrated the supervised training of a ML model capable of assessing spectra quality using a quality score label. That the ML model surpasses analytical performance in this assessment proves that the \textit{aurelia} synthetic spectra are a suitable proxy for experimental data in the training of ML algorithms. By combining this tool with an autonomously driven spatial motor~\cite{Agustsson2024}, one can automate the alignment and assessment of any sample -- the most tasking phase of any ARPES experiment. Such tools have already been created and tested for various scanning probe techniques~\cite{Krull2020}, though the applications are typically limited to one or a few samples. 

The strong performance of the deep learning model stems from its ability to learn directly from a wide distribution of data, rather than relying on fixed heuristics. While the quality score labels are generated by a rule-based algorithm, they are intended to approximate human assessments of spectral quality — a concept that is inherently subjective and difficult to define rigorously. Deep neural networks can capture subtle, high-dimensional patterns across the entire image. This allows the model to learn nuanced distinctions between spectra that go beyond what can be encoded in handcrafted metrics. Additionally, the use of dynamically generated training data enhances the model's ability to generalize. With 50,000 new samples per epoch, drawn from a wide range of simulation conditions, the model encounters varying intensities, noise levels, and feature sharpness — reflecting the diversity seen in real experiments. While the simulation-based quality scores serve as an initial guide, they are not the absolute ground truth. In practice, the deep learning model can yield assessments that better reflect experimental intuition, even when trained on these approximated labels.

While we have trained the model in quality assessment, the simulations are generalizable to train ML models in various tasks. For instance, one can train a model to recognize flake domains with large angular displacement ($\theta_0$, $\phi_0$) to better select uniform regions of the sample to measure. Extending this to rotational domains ($\alpha_0$), the ML model can accurately determine the relative angle in twisted heterostructures. One can also train an algorithm to accurately extract energy gaps under temperature and resolution broadening, which can be used to find superconducting gap inhomogeneities or pump-induced changes in a time-resolved ARPES experiment. To this end, the code for the simulation is open access and freely adaptable~\cite{NaGit2024}. As ARPES experiments become ever more complex, ML tools will be instrumental in taking over labour-intensive and time-consuming tasks such as quality assessment, denoising~\cite{Restrepo2022}, feature extraction\cite{Peng2020}, analysis~\cite{Ekahana2023, Mortensen2023, majchrzak2025}, and classification. While the scientific community prepares the infrastructure necessary to store and share data and metadata, synthetically generated data are an effective alternative for training ML algorithms. 

\section{Acknowledgements}
We would like to gratefully acknowledge Peter C. Moen, Bradley G. Guislain, Giorgio Levy, Sergey Zhdanovich and Jerry Dadap for the supporting discussions. We also acknowledge Dr. Xin Lu for their support and for sharing valuable code snippets used in the machine-learning pipeline.
This research was undertaken thanks in part to funding from the Max Planck-UBC-UTokyo Centre for Quantum Materials and the Canada First Research Excellence Fund, Quantum Materials and Future Technologies Program. This project is also funded by the Natural Sciences and Engineering Research Council of Canada (NSERC); Canada Foundation for Innovation (CFI); the British Columbia Knowledge Development Fund (BCKDF); the Department of National Defence (DND); the Gordon and Betty Moore Foundation’s EPiQS Initiative, Grant GMBF4779 (A.D.); the Canada Research Chairs Program (A.D.); and the CIFAR Quantum Materials Program (A.D.).

\bibliographystyle{unsrt}
\bibliography{manuscript}

\end{document}